\documentstyle[12pt,oneside,amssymb,array,amstex]{amsart}

\newcommand{\al}{\alpha}
\newcommand{\be}{\beta}
\newcommand{\pa}{\partial}

\newcommand{\la}{\lambda}

\newcommand{\ga}{\gamma}
\newcommand{\om}{\omega}
\newcommand{\de}{\delta}

\newcommand{\rar}{\rightarrow}

\def\fun#1#2{\lower3.6pt\vbox{\baselineskip0pt\lineskip.9pt
  \ialign{$\mathsurround=0pt#1\hfil##\hfil$\crcr#2\crcr\sim\crcr}}}
\newskip\humongous \humongous=0pt plus 1000pt minus 1000pt

\newif\ifdtup

\begin{document}

\begin{titlepage}

\begin{flushright}
M\'exico ICN-UNAM 05/00\\ April 16, 2000
\end{flushright}

\vskip 1.6cm

\begin{center}

{\Large Canonical Discretization \\[15pt]
 I. Discrete faces of (an)harmonic oscillator}

\vskip 0.6cm

{\it Alexander Turbiner}$^{\dagger}$
\vskip 0.5cm
Instituto de Ciencias Nucleares, UNAM, Apartado Postal 70--543,
\\04510 Mexico D.F., Mexico

\end{center}

\vskip 2.cm

\centerline{Abstract}

\begin{quote}
A certain notion of canonical equivalence in quantum mechanics is
proposed. It is used to relate quantal systems with discrete ones.
Discrete systems canonically equivalent to the celebrated harmonic
oscillator as well as the quartic and the quasi-exactly-solvable
anharmonic oscillators are found. They can be viewed as a
translation-covariant discretization of the (an)harmonic
oscillator preserving isospectrality. The notion of the
$q-$deformation of the canonical equivalence leading to a
dilatation-covariant discretization preserving polynomiality of
eigenfunctions is also presented.
\end{quote}
\vskip 2.5cm

\vfill
\noindent
$^\dagger$On leave of absence from the Institute for Theoretical
and Experimental Physics, Moscow 117259, Russia\\ E-mail:
turbiner@@xochitl.nuclecu.unam.mx

\end{titlepage}
Two classical mechanical systems related by a canonical
transformation are equivalent (for general discussion see, for
example, \cite{llcm}). It is quite natural to try to adopt a
similar notion for the quantum mechanical systems making a
difference that classical canonical transformations should be
replaced by their quantum counterpart, which means that the
Poisson bracket is replaced to the Lie bracket, $[P(p,q),
Q(p,q)]=1$. Present work is an attempt to explore this natural
definition taking as an example both (an)harmonic oscillator and a
certain particular type of quantum canonical transformations.
There are well-known difficulties on this way and we will try to
indicate and overtake them. Meanwhile, we will call systems
related through a quantum canonical transformation {\it
canonically equivalent}.

Let us remind that any quantum-mechanical system with a
Hamiltonian ${\cal H}(x,\hat p)$ is intrinsically related to a
Heisenberg algebra $[x,\hat p]=-i$. Hamiltonian can be treated as
an operator (or, an element of the Heisenberg-Weyl algebra) acting
in the quantum-mechanical phase space which is defined as an
object consisting of the universal enveloping Heisenberg algebra
(the Heisenberg-Weyl algebra) and the vacuum $|0>$, such that
$\hat p |0>=0$. From this point of view, canonical transformations
are nothing but changes of variables in the phase space preserving
polarization or, in other words, preserving the property that the
Heisenberg algebra is the underlying algebra of the system. This
construction can be generalized by considering formally the
Hamiltonian as an operator acting in the $q$-deformed phase space.
It implies the $q$-deformed Heisenberg algebra as the underlying
algebra, $x {\hat p} - q{\hat p} x = -i$ of the system
 \footnote{It also allows a possible modification of the
 Hamiltonian by inserting the parameter $q$ in appropriate places},
and taking a class of transformations preserving this algebraic
structure. It leads naturally to the Fock space formalism as an
adequate language of our study.

The harmonic oscillator as well as anharmonic, both classical and
quantum plays a fundamental role in physics. A particular goal of
present study is to find discrete systems those can be related to
quantum-mechanical (an)harmonic oscillator via a quantum canonical
transformation. It will be considered two types of discrete
systems: translation-covariant (uniform grid) (i) and
dilatation-covariant (exponential grid) (ii). In what follows it
will be used a Fock space formalism. In the next paper the case of
Coulomb, Poschl\"e-Teller and Morse potentials will be presented
\cite{tII}.

Important general remark should be made in row. Generically, if
two systems I and II are related by a gauge transformation and/or
by a change of variables as equivalent, namely, $${\cal H}_I (x)=
g(y)^{(-1)} {\cal H}_{II}(y)g(y)|_{y=y(x)}$$ these systems will be
treated as equivalent.

\section{\it Harmonic oscillator in Fock space}

The Hamiltonian of harmonic oscillator is defined by
\begin{equation}
\label{e1.1}
        {\cal H}= -\frac{1}{2} \frac{\pa^{2}}{\pa x^{2}} +
       \frac{ \om^2}{2} x^{2} \ ,
\end{equation}
where $\om$ is the oscillator frequency. The configuration space
of (1) is the whole real line, $x \in (-\infty,\infty)$. The
eigenfunctions and eigenvalues are given by
\begin{equation}
\label{e1.2}
\Psi_{k}(x) = H_k(\sqrt{\om}x) e^{-\om\frac{x^{2}}{2}}\ ,
\ E_k=\om (k +1/2),\ k=0,1,\ldots
\end{equation}
where $H_k(y)$ is the $k$th Hermite polynomial in standard
notation. Without a loss of generality we put all normalization
constants in (2) equal to 1. The Hamiltonian (\ref{e1.1}) is
$Z_2$-invariant, $x \rar -x$, which leads to two families of
eigenstates: even and odd, or, in other words, symmetric and
anti-symmetric with respect of reflection, correspondingly. This
property is coded in parity of the Hermite polynomials and is
revealed by the relation:
\begin{equation}
\label{e1.3}
 H_{2n+p}(\sqrt{\om}x)= \om^{\frac{p}{2}} x^p L_n^{(p-\frac{1}{2})}(\om x^{2})\ ,\ n=0,1,\ldots
\end{equation}
where $L_n^{(\al)}(y)$ is the $n$th associated Laguerre
polynomial in standard notation, and the parameter $p=0,1$
determines parity, $P=(-1)^p$. Hereafter we can call
\begin{equation}
\label{e1.4}
\Psi_{0}^{(p)}(x) = x^p e^{-\om\frac{x^{2}}{2}}\ ,
\end{equation}
the ground state (the lowest energy state) of parity $P$. Thus,
the formula (4) makes an unification of both possible values of
parity and for the sake of simplicity we will call (4) the ground
state eigenfunction without specifying parity.

Let us make a gauge rotation of the Hamiltonian (1) taking as a
gauge factor the ground state eigenfunction (4) and changing the
variable $x$ to $y=\om x^2$, which incorporate the reflection
symmetry. After dropping out non-essential constant term,
$(\frac{1}{2}+p)$, which cause a change the reference point for
energy only, we get an operator
\begin{equation}
\label{e1.5}
 h(y,\pa_y)\ =\ \frac{1}{\om}\ (\Psi_0^{(p)}(x))^{-1}{\cal
 H}\Psi_0^{(p)}(x)\mid_{y=\om x^2}\ =\ -2y\pa_y^2 +
 2(y-p-\frac{1}{2})\pa_y\
\end{equation}
with the spectrum $E_n=2n$, where $n=0,1,2\ldots$
 \footnote{It is necessary to emphasize that the spectrum is
 defined by action of the operator $2y\pa_y$ in (5) on monomial:
 $(2y\pa_y) y^n=2n y^n$. It is the only operator of
 zero-grading in (5)}.
Now the eiegenvalue problem for the operator (\ref{e1.5}) is
defined on the half-line $y \in [0,\infty)$. The operator (5)
simultaneously describes a family of eigenstates of positive
parity if $p=0$ and a family of eigenstates of negative parity if
$p=1$. We will call (5) {\it the algebraic form} of the
Hamiltonian of the harmonic oscillator. The word `algebraic'
reflects the fact that the operator (5) has a form of linear
differential operator with polynomial coefficients and
furthermore possesses infinitely-many polynomial eigenfunctions.
The latter implies that any eigenfunction can be found by
algebraic means by solving a system of linear algebraic
equations.

The algebraic form (5) admits a generalization of the original
Hamiltonian (1) we started with. If we assume that the parameter
$p$ can take any real value, $p>-1/2$, one can make an inverse
gauge transformation of the operator (5) back to the Hamiltonian
form and we arrive at
\[
{\om}\ y^{p/2}e^{- y/2}\bigg[-2y\pa_y^2 + 2(
y-p-\frac{1}{2})\pa_y\bigg]y^{-p/2}e^{y/2}\mid_{x=\sqrt{
\frac{y}{\om}}}
\]
\begin{equation}
\label{e.1.6}
 =\bigg[-\frac{1}{2}\pa_x^2 + \frac{\om^2}{2} x^2
 +\frac{p(p-1)}{2x^2}\bigg]\ \equiv \ {\cal H}_k\ ,
\end{equation}
which is known in literature as Kratzer Hamiltonian. It is worth
to mention that this Hamiltonian coincides with 2-body Calogero
Hamiltonian. Also this Hamiltonian appears as a Hamiltonian of
the radial motion of multidimensional spherical-symmetrical
harmonic oscillator.  Hereafter we will continue to call the
system characterized by the Hamiltonian (6) {\it the harmonic
oscillator}.

The resulting Hamiltonian (6) is characterized by the
eigenfunctions
\begin{equation}
\label{e1.7}
\Psi_{k}(x) =  x^p L_n^{(p-\frac{1}{2})}(\om x^{2})
e^{-\om\frac{x^{2}}{2}}\ ,
\end{equation}
which coincides with (2) at $p=0,1$. The spectrum (6) is still
equidistant with energy gap $\om$ and after appropriate shift of
the reference point it coincides with the spectrum of the
original harmonic oscillator (1). Thus, the deformation of (1) to
(6) is isospectral which is, of course, well-known.

In order to move ahead let us introduce a notion of the Fock
space. It will be a natural formalism to study canonical
transformations. Take two operators $a$ and $b$ obeying the
commutation relation
\begin{equation}
\label{e.2.1}
             [a,b] \equiv ab  -  ba \ =\ I,
\end{equation}
with the identity operator $I$ on the r.h.s. -- they span a
three-dimensional Lie algebra which is called the Heisenberg
algebra $h_3$. By definition the universal enveloping algebra of
$h_3$ is the algebra of all normal-ordered polynomials in $a,b$:
any monomial is taken to be of the form $b^k a^m$
\footnote{Sometimes this is called the Heisenberg-Weyl algebra}.
If, besides the polynomials, all entire functions in $a,b$ are
considered, then the {\it extended} universal enveloping algebra
of the Heisenberg algebra appears or in other words, the extended
Heisenberg-Weyl algebra. In the (extended) Heisenberg-Weyl
algebra one can find the non-trivial embedding of the Heisenberg
algebra: non-trivial elements obeying the commutation relations
(\ref{e.2.1}), whose can be treated as a certain type of quantum
canonical transformations. We say that the {\it (extended) Fock
space, $\cal F$} is determined if we take the (extended)
universal enveloping algebra of the Heisenberg algebra and attach
to it the vacuum state $|0>$ defined as
\begin{equation}
\label{e.2.2}
a|0>\  = \ 0\ .
\end{equation}
It is easy to check that the following statement holds. If the
operators $a,b$ obey the commutation relation (\ref{e.2.1}), then
the operators
\[
J^+_n = b^2 a - n b\ ,
\]
\begin{equation}
\label{e.2.3}
J^0_n = ba - {n \over 2}\ ,
\end{equation}
\[
J^-_n=a\ ,
\]
span the $sl_2$-algebra with the commutation relations:
\[
[J^0,J^{\pm}]=\pm J^{\pm}\ ,\  [J^+,J^-]=-2J^0\ ,
\]
where $n \in {\bf C}$
\footnote{For details and discussion see, for example,
\cite{Turbiner:1997}}. For the realization (\ref{e.2.3}) the
quadratic Casimir operator is equal to
\begin{equation}
\label{e.2.4}
C_2 \equiv \frac{1}{2}\{J^+_n,J^-_n\} - J^0_n J^0_n =
-\frac{n}{2} \bigg(\frac{n}{2} + 1\bigg)\ ,
\end{equation}
where $\{,\}$ denotes the anticommutator and is $c-$number. If $n
\in \mathbb{Z}_+$, then (\ref{e.2.3}) possesses a finite-dimensional,
irreducible representation in Fock space leaving invariant the
linear space of polynomials in $b$ acting on vacuum:
\begin{equation}
\label{e.2.5}
{\cal P}_{n}(b) \ = \ \langle 1, b, b^2, \dots , b^n \rangle
|0\rangle,
\end{equation}
of dimension $\dim{\cal P}_{n}=(n+1)$. It is evident that any
operator which is a polynomial in generators $J^{+,0,-}_n$
preserves the space ${\mathcal P}_{n}(b)$ (and converse is also
correct \cite{Turbiner:1994}). Such an operator we call {\it
$sl_2$-quasi-exactly-solvable operator}. Thus, the most general
$sl_2$-quasi-exactly-solvable operator in the Fock space having a
form of a polynomial in $a$ of degree not higher than two is given
by
\[
 T_2 =  c_{++} J^+_n J^+_n  + c_{+0} J^+_n  J^0_n  + c_{00}
 J^0_n  J^0_n  + c_{0-} J^0_n  J^-_n  + c_{--} J^-_n  J^-_n  +
\]
\begin{equation}
\label{e.2.5.1}
 c_+ J^+_n  + c_0 J^0_n  + c_- J^-_n  + c\ ,
\end{equation}
where $c_{\al \beta}, c_{\al}, c$ are arbitrary c-numbers, or
after the substitution (\ref{e.2.3}) in explicit form
\begin{equation}
\label{e.2.5.2}
 T_2(b,a)\ =\ - P_{4}(b) a^2 \ +\ P_{3}(b) a \ +\ P_{2}(b) ,
\end{equation}
where the $P_{j}(b)$ are polynomials of $j$th order with
coefficients related to  $ c_{\al \beta}, c_{\al}, c$ and $n$.

The spaces ${\cal P}_n$ possess a property that ${\cal P}_n
\subset {\cal P}_{n+1}$ for each $n \in \mathbb{Z}_+$ and form an
infinite flag (filtration) and $$\bigcup_{n \in \mathbb{Z}_+}
{\cal P}_n = \mathcal{P}.$$ Hence it is evident that any operator
which is a polynomial in generators $J^{0,-}_n$ only preserves the
flag of ${\mathcal P}$ \cite{Turbiner:1994}. Such an operator we
call {\it $sl_2$-exactly-solvable operator}. Thus, the most
general $sl_2$-exactly-solvable operator in the Fock space having
a form of a polynomial in $a$ of degree not higher than two is
given by
\begin{equation}
\label{e.2.5.3}
 E_2 =   c_{00} J^0  J^0  + c_{0-} J^0  J^-
 + c_{--} J^-  J^-  +
 c_0 J^0  + c_- J^-  + c\ ,
\end{equation}
where $J^{\pm,0}\equiv J^{\pm,0}_0$ and $c_{\al \beta}, c_{\al},
c$ are arbitrary c-numbers. After the substitution (\ref{e.2.3})
in explicit form
\begin{equation}
\label{e.2.5.4}
 E_2(b,a)\ =\ -Q_{2}(b) a^2 \ +\ Q_{1}(b) a \ +\ Q_{0}\ ,
\end{equation}
where the $Q_{j}(b)$ are polynomials of $j$th order with
coefficients related to  $ c_{\al \beta}, c_{\al}, c$.

Now we can introduce a notion of the spectral problem in the Fock
space. Let $L(b,a)$ is an element of the Heisenberg-Weyl algebra.
By definition, to solve the spectral problem for the operator
$L(b,a)$ is to find a set of the elements $\{\phi (b)\}$ in the
Heisenberg-Weyl algebra and a corresponding set of parameters
$\{\la \}$ for those the equation
\begin{equation*}
L(b,a)\phi (b)|0>\ =\ \la \phi (b) |0> \qquad \qquad \qquad (*)
\end{equation*}
is fulfilled. We will call $\{\phi (b)\}$ and $\{\la \}$ the
eigenfunctions and eigenvalues, respectively. Attempting to study
the spectral problem $(*)$ immediately leads to a delicate
question about convergence of the operator series. In order to
avoid possible difficulties we will restrict our consideration by
the cases when the polynomial in $b$ eigenfunctions appear.

\section{\it Translation-covariant discretization}

Take as an example the $sl_2$-exactly-solvable operator of the
following form
\begin{equation}
\label{e.2.6}
 h^{(1)}(b,a)\ =\ -2J^0 J^- +2 J^0 -2 (p+\frac{1}{2}) J^-
 \ =\ -2ba^2 +2(b-p-\frac{1}{2})a\ ,
\end{equation}
where $p$ is a parameter and $J^{\pm,0}\equiv J^{\pm,0}_0$ (see
(10)). Simple analysis leads to a statement that the
eigenfunctions of $h_f$ are the associated Laguerre polynomials of
the argument $b$, $L_n^{(p-\frac{1}{2})} (b)$ and their
eigenvalues, $E_n=2n,\ n=0,1,2,\ldots$
 \footnote{As in (5) the eigenvalues are defined by action of the
 operator $(2ba)$ in (13) on monomials:
 $(2ba) b^n= 2n b^n + 2b^{n+1} a$. The second term disappears
 after action on the vacuum}.

As the next step we consider two different realization of the
Heisenberg algebra (\ref{e.2.1}) in terms of differential and
finite-difference operators. A traditional realization of
(\ref{e.2.1}) appearing in text-books is the so-called
coordinate-momentum representation:
\begin{equation}
\label{e.3.1}
a\ =\ \frac{d}{dy} \equiv \pa_y\ ,\ b\ =\ y\ ,
\end{equation}
(see, for example, the book by Landau and Lifschitz \cite{llqm}),
where the operator $b=y$ stands for the multiplication operator :
$b f(y)= y f(y)$. In this case the vacuum is a constant and
without a loss of generality we put $|0>\  = \ 1$. However, there
exists another realization of (\ref{e.2.1}) in terms of
finite-difference operators. It is a finite-difference analogue of
(\ref{e.3.1}):
\begin{equation}
\label{e.3.2}
a\ =\ {\cal D}_+ ,\quad b\ =\ (y+\al)(1-\de{\cal D}_-) \ ,
\end{equation}
where
\[
{\cal D}_{\pm} f(y) = \frac{f(y\pm\de) - f(y)}{\pm\de}\ ,
\]
(cf.\cite{Smirnov:1995}) is the finite-difference operator. It
represents what can be called a {\it $\de-$discretization} of the
derivative: ${\cal D}_{\pm} \rar \pa_y$, if $\de \rar 0$ and,
thus, which can be called $\de-$derivative
 \footnote{This finite-difference operator is also known in
 mathematics literature as the N\"orlund derivative (see, for
 example,\cite{finite}) while we prefer to use a name
 {\it $\de$-derivative} in order to distinguish from
 {\it $q$-derivative} (see below) -- another type of
 discretization}. In general, $\de,\al$ can be any
complex numbers and ${\cal D}_{\pm}(-\de) = {\cal D}_{\mp}(\de)$.
It is necessary to emphasize that from physical point of view the
operator $b$ in (\ref{e.3.2}) is nothing but the
canonical-conjugate to a discrete momentum operator defined by
$\de-$derivative. So, (\ref{e.3.2}) represents two-parametric
family of quantum canonical transformations. For $\al=0$ and $\de
\rar 0$, the formulas (\ref{e.3.2}) become (\ref{e.3.1}). It
allows to interpret (\ref{e.3.2}) as the continuous deformation of
(\ref{e.3.1}).

A remarkable property of this realization is that the vacuum can
be taken as a constant
 \footnote{Any $\de-$periodic function can be chosen as a vacuum
 for (\ref{e.3.2}) as well }
and thus, without loss of generality, it can be placed as $|0>\ =\
1$ for both cases (\ref{e.3.1})-(\ref{e.3.2}). Realization
(\ref{e.3.2}) is translation-covariant: under a linear shift of
variable $y \rar y+B$ the functional form of (\ref{e.3.2}) is
preserved.

Substitution of (\ref{e.3.1}) into (\ref{e.2.6}) leads to the
operator (5) -- the algebraic form of the Hamiltonian of the
harmonic oscillator. Thus, the operator (\ref{e.2.6}) can be
called the {\it the Hamiltonian of the harmonic oscillator in the
Fock space}. It is evident that the procedure of realization of
the Heisenberg generators $a,b$ by concrete operators
(differential, finite-difference, discrete) provided that the
vacuum remains unchanged leaves any polynomial operator in $a,b$
isospectral.

Now let us study another `face' of harmonic oscillator by
substituting the realization of the Heisenberg algebra by
finite-difference operators (\ref{e.3.2}) into (\ref{e.2.6}). It
results to
\begin{equation}
\label{e.3.3}
 h_{\de}^{(1)} (y, D_{\pm})=
 -\frac{2}{\de}[y+\al+\de(p+\frac{1}{2})]D_+ + 2(1 +\frac{1}
 {\de})(y+\al)D_- \ .
\end{equation}
In this realization the corresponding spectral problem $(*)$ can
be written as
\[
-\frac{2}{\de^2}\big[y+\al+\de(p+\frac{1}{2})\big]\phi(y+\de)
+
\frac{2}{\de}\big[(1+\frac{2}{\de})y+\al+p+\frac{1}{2}\big]\phi(y)
\]
\begin{equation}
\label{e.3.4}
- \frac{2}{\de}(1+\frac{1}{\de})(y+\al)\phi(y-\de)
= E \phi(y) \ .
\end{equation}
It is worth to note that the Askey condition that the sum of the
coefficients in front of unknown functions should be zero is
fulfilled at $E=0$. It is equivalent to a statement that the
equation (\ref{e.3.4}) possess an eigenfunction which is a
constant.

In general, the operator $h_{\de}(y, D_{\pm})$ is a non-local,
three-point, finite-difference and translation-covariant operator.
This operator is canonically-equivalent to the harmonic
oscillator. Pictorially it is illustrated by Fig.1.
 \vskip .5cm
 \noindent
 \unitlength.8pt
\begin{picture}(400,50)(-10,-20)
 \linethickness{1.2pt}
 \put(60,10){\line(1,0){250}}
 \put(120,10){\circle*{5}}
 \put(180,10){\circle*{5}}
 \put(240,10){\circle*{5}}
 \put(90,-15){$\phi(y-\de)$}
 \put(170,-15){$\phi(y)$}
 \put(230,-15){$\phi(y+\de)$}
\end{picture}
\begin{center}
Fig. 1. \ Graphical representation of the operator (\ref{e.3.3})
\end{center}

So, in the $y-$space we have the uniform grid, linear lattice.
However, from the point of view of the original configuration
space, $x-$space where the harmonic oscillator (1) is defined we
have a square lattice.

In order to find the eigenfunctions of (\ref{e.3.3}) a certain
trick can be used. One can easily show that the following equality
holds
\[
[ (y+\al)e^{-\de \pa_y}]^n {\it I}\ =\ (y+\al)^{(n)} {\it I}\ ,
\]
where $(y+\al)^{(n+1)}=(y+\al)(y+\al-\de)\ldots (y+\al-n\de)$ is a
so-called {\it quasi-monomial} or {\it generalized monomial} and
$I$ is the identity operator. Then making use this relation it is
not difficult to prove that the eigenfunctions of (\ref{e.3.3})
remain polynomials in $y$ and, furthermore, they can be found
explicitly in rather elegant way
\begin{equation}
\label{e.3.5}
\phi_n (y)\ =\ {\hat L}_n^{(p-\frac{1}{2})}(y,\de)=\sum_{\ell=0}^n
 a_{\ell}^{(p-\frac{1}{2})} (y+\al)^{(\ell)}\ ,
\end{equation}
where $a_{\ell}^{(p-\frac{1}{2})}$ are the coefficients in the
expansion of the Laguerre polynomials, $L_n^{(p-\frac{1}{2})}(y)=
\sum_{\ell=0}^n a_{\ell}^{(p-\frac{1}{2})} y^{\ell}$. We call
these polynomials the {\it modified associated Laguerre
polynomials}. Simultaneously, the eigenvalues in the equation
(\ref{e.3.4}) remain equal to $(2n), n=0,1,2\ldots$ and they are
the same as the eigenvalues of the harmonic oscillator problem
(1) as well as (6) and (13). Thus, one can say that the operator
(\ref{e.3.3}) defines a {\it finite-difference or $\de-$discrete
algebraic form of harmonic oscillator} Hamiltonian. Without loss
of generality one can place $\al=0$ in (\ref{e.3.3}),
(\ref{e.3.4}), (\ref{e.3.5}).

There exists two non-trivial particular cases of (\ref{e.3.3}).
The first case corresponds to the spacing $\de=-1$. It leads to a
disappearance of the term proportional to $D_-$ and, thus, the
operator (\ref{e.3.3}) becomes two-point operator (!)
\[
 h_{\de}^{(1)}(y, D_{\pm})= 2[y+\al-(p+\frac{1}{2})]D_+|_{\de=-1}
\equiv
\]
\begin{equation}
\label{e.3.3.1}
 2[y+\al-(p+\frac{1}{2})]D_-|_{\de=1} \ .
\end{equation}
It is worth noting that the spectrum in this case is defined by
the term $(2yD_+|_{\de=-1})$ stemming from the negative grading
term $(2ba^2)$ in (\ref{e.2.6}). Breaking a condition of
performing the canonical transformations we re-insert the
parameter $\de$ in (\ref{e.3.3.1}) in naive, straightforward
manner: $D_+|_{\de=-1} \rar D_+(\de)$. It results to the operator
\begin{equation}
\label{e.3.3.2}
 h_{\de}^{(1)}(y, D_{\pm})\ =\ 2[y+\al-(p+\frac{1}{2})]D_+  \ ,
\end{equation}
which is isospectral to (5), (\ref{e.2.6}) for any $\de$. In
limit $\de \rar 0$ it leads to the first-order differential
operator
\begin{equation}
\label{e.3.3.3}
 h_{\de}^{(1)}(y,\pa_y)\ =\ 2[y+\al-(p+\frac{1}{2})]\pa_y  \ .
\end{equation}
Although, the operators (\ref{e.3.3.1}), (\ref{e.3.3.2}) are
isospectral to various forms of the harmonic oscillator (1), (6),
those operators are not related to each other by a quantum
canonical transformation and therefore are not canonically
equivalent. Above-mentioned isospectral transition from the
second-order differential operator (5) to the first order
differential operator (\ref{e.3.3.3}) or from finite-difference
one (\ref{e.3.3}) to (\ref{e.3.3.2}) reminds celebrated Bargmann
transformation (see for example \cite{perelomo}). It is necessary
to emphasized that the operator (\ref{e.3.3.2}) has
infinitely-many polynomial eigenfunctions, however, unlike the
operators (13),(16), the operator (20) is not
$sl_2$-exactly-solvable (!). Also it can not be rewritten in terms
of the generators $a,b$ of the Heisenberg algebra (8) and does not
belong to the Fock space (see discussion below).

Another important particular case occurs if the spacing $\de=-2$.
In this case the operator (\ref{e.3.3}) again becomes two-point
one
\begin{equation}
\label{e.3.4.1}
-\frac{1}{2}\big[y+\al-2(p+\frac{1}{2})\big]\phi(y-2)
+\frac{1}{2}(y+\al)\phi(y+2)
= \tilde E \phi(y) \ ,
\end{equation}
where a new spectral parameter $\tilde E = E +
\al+p+\frac{1}{2}$.

\begin{quote}
{\it Remark.} The function $\phi(y)$ in the r.h.s. of
(\ref{e.3.4}) can be replaced by $\phi(y+\de)$ or $\phi(y-\de)$,
or by a linear combination of $\phi(y\pm\de), \phi(y)$. It does
not change the statement that (\ref{e.3.4}) has infinitely-many
polynomial eigenfunctions. This procedure preserves isospectral
property. However, such changes of the r.h.s. lead to a
replacement of the original standard spectral problem $(*)$ by a
`generalized' spectral problem. In this case the r.h.s. contains
an operator other than the operator of multiplication on a
function. In general, a physical relevance of such
right-hand-sides is unclear. It is worth to note that it reminds
the Sturm representation of the Coulomb problem when the energy is
kept fixed but a set of discrete electric charges for which a
corresponding excited state energy is equal to this energy is
studied. In this formulation the r.h.s. of the Coulomb problem
contains the Coulombic interaction potential as the weight factor.
\end{quote}

A natural question can be posed about the most general
second-order linear differential operator, which (i) is
isospectral to the harmonic oscillator (1),(6); (ii) has
infinitely-many polynomial eigenfunctions and (iii) related to (5)
by a canonical transformation. Following the Theorem
\cite{Turbiner:1994} one can derive the most general operator with
above properties
\[
h_{\de,g}(y,\pa_y)\ =\ -2 (AJ^0+BJ^-) J^- +2 J^0 -2 C J^-\ =
\]
\begin{equation}
\label{e4.1}
-2(Ay+B)\pa_y^2 + 2 (y-C)\pa_y\
\end{equation}
where $A,B,C$ are arbitrary constants and the generators
(\ref{e.2.3}) are realized by differential operators
(\ref{e.3.1}). However, by a linear change of variable, $y \rar
\al y + \be$ with appropriate $\al \neq 0,\be$ the operator
(\ref{e4.1}) can be transformed into (5). It reflects a fact that
the linear space of polynomials ${\cal P}_n(x)$ is invariant under
a linear transformation: $x \rar \ga_1 x+\ga_2, \ga_1 \neq 0$.
Thus, without loss of generality we can put $A=\al=1$ and also
$C=p+1/2$. The eigenfunctions of (\ref{e4.1}) remain the Laguerre
polynomials but of a shifted argument, $L_n^{(p-\frac{1}{2})} (y
+\be)$. It leads to a statement that among the second-order
differential operators there exists no non-trivial isospectral
deformation of the harmonic oscillator potential preserving
polynomiality of the eigenfunctions.

The operator (\ref{e4.1}) can be rewritten in the Fock space
formalism by using (\ref{e.3.1})
\begin{equation}
\label{e4.2}
h_g(b,a)\ =\ - 2(b+B)a^2 + 2(b-C)a\ .
\end{equation}
It is evident that the operator (\ref{e4.2}) is the most general
second order polynomial in $a$, which is isospectral to
(\ref{e.2.6}) and also preserves the space of polynomials
(\ref{e.2.5}). By substitution (\ref{e.3.2}) into the operator
(\ref{e4.2}) it becomes a finite-difference operator
\begin{equation}
\label{e.4.3}
h_{\de,g}(y,D_{\pm})\ =\ -2BD_+^{2}-
\frac{2}{\de}[y+\al+\de C]D_+ + 2(1 +\frac{1}{\de})(y+\al) D_-
\end{equation}
(cf. (\ref{e.3.3})). Thus, in the realization by
finite-difference operators the corresponding spectral problem
(*) has the form
\[
\frac{-2B}{\de^2}\phi(y+2\de)-
\frac{2}{\de^2}\big[y+\al-2B +\de C\big]\phi(y+\de)
\]
\begin{equation}
\label{e.4.4}
 + \frac{2}{\de}\big[(1+\frac{2}{\de})(y+\al)-\frac{B}{\de}+
 C\big]\phi(y)
 - \frac{2}{\de}(1+\frac{1}{\de})(y+\al)\phi(y-\de)\ =\ E \phi(y)\ ,
\end{equation}
and is characterized by existence of infinitely family of
eigenfunctions given by polynomials in $y$.

The operator $h_{\de}(y,D_{\pm})$ now becomes the four-point
finite-difference operator, see Fig.2 .
\vskip .5cm
\noindent
\unitlength.8pt
\begin{picture}(400,50)(-10,-20)
\linethickness{1.2pt}
\put(60,10){\line(1,0){300}}
\put(120,10){\circle*{5}}
\put(180,10){\circle*{5}}
\put(240,10){\circle*{5}}
\put(300,10){\circle*{5}}
\put(90,-15){$\phi(y-\de)$}
\put(170,-15){$\phi(y)$}
\put(220,-15){$\phi(y+\de)$}
\put(290,-15){$\phi(y+2\de)$}
\end{picture}
\begin{center}
Fig. 2. \ Graphical representation of the operator (\ref{e.4.3}).
\end{center}

It is quite surprising that a simple transformation like linear
shift of variable $y$ in differential operator (\ref{e.3.3})
(which leads to nothing non-trivial, see above) leads to
occurrence of the extra point in the isospectral finite-difference
counterpart changing a type of its non-locality. Nevertheless,
these operators remain to be canonically equivalent being related
each other by a canonical transformation. So, by a canonical
transformation one can change a non-local nature of
finite-difference operators.

It is quite interesting to abandon the condition of canonical
equivalence and pose a question about the most general
differential (finite-difference) operators isospectral to the
harmonic oscillator and possessing the infinitely-many polynomial
eigenfunctions. One can easily show that ignoring the condition
of canonical equivalence leads to nothing new for differential
operators. However, for the case of finite-difference operators
much wider class of operators occurs than (\ref{e.3.3}) or
(\ref{e.4.3}). As a natural constraint we have to impose a
condition of maximal number of points in the finite-difference
operators we search for. Let us consider three-point operators
(see Fig.1). A simple analysis leads to a statement that the
three-parametric operator
\begin{equation}
\label{e.4.5}
 {\tilde h}_{\de} (y, D_{\pm})= 2[(1-A)y+C_+)]D_+ + 2[Ay+C_-)D_-
 \ ,
\end{equation}
where $A,C_{\pm}$ are parameters, is the most general three-point
finite-difference operator with infinitely-many polynomial
eigenfunctions, which is isospectral to the harmonic oscillator.
Of course, the operators (16) and (18) are particular cases of
(\ref{e.4.5}). The form (\ref{e.4.5}) can be understood as a
consequence of the Theorem which states that {\it any
finite-difference operator $h_{\de}(y, D_{\pm})$ preserving the
infinite flag of polynomials
\begin{equation}
\label{e.4.6}
 {\cal P}_{n}(y) \ = \ \langle 1, y, y^2, \dots , y^n
 \rangle \ ,
\end{equation}
should have a representation in terms of the operators:}
\begin{equation}
\label{e.4.7} J^-_{\pm}\ =\ D_{\pm}\ ,\ J^0_{\pm}\ =\ yD_{\pm}\ .
\end{equation}
As a remark we should note that the generators (\ref{e.4.7}) do
not span an algebra closed with respect to commutators. However,
if we consider a linear space spanned by (\ref{e.4.7}), then the
Heisenberg algebra (\ref{e.3.1}) appears as its subspace.

It is worth to mention that the canonical discretization of the
harmonic oscillator can be made directly in the configuration
$x-$space where the original harmonic oscillator (1) is defined.
If we make a gauge transformation (5) with (4) at $p=0$ as the
gauge factor but without the change of variable, then
\begin{equation}
\label{e.4.8}
 h^{(2)} (x,\pa_x)\ =\ \frac{1}{\om}(\Psi_0^{(0)}(x))^{-1}{\cal
 H}\Psi_0^{(0)}(x)\ =\ -\frac{1}{2\om} \pa_x^2 + x\pa_x\ ,
\end{equation}
is another algebraic form of the harmonic oscillator (cf.(5)). The
operator (\ref{e.4.8}) has the Hermite polynomials,
$H_k(\sqrt{\om}x)$, as the eigenfunctions. Now we can make
$\de-$discretization firstly rewriting (\ref{e.4.8}) in the Fock
space formalism
\begin{equation}
\label{e.4.9}
 h^{(2)}(b,a)\ =\ -\frac{1}{2\om} a^2 + ba\ =
 \ -\frac{1}{2\om}J^- J^- + J^0  \ ,
\end{equation}
(cf.(\ref{e.2.6})) and then realizing $a,b$ by finite-difference
operators (\ref{e.3.2})
\begin{equation}
\label{e.4.10}
 h_{\de}^{(2)}(x,D_{\pm})\ =\ -\frac{1}{2\om}D_+^{2} +
 (x+\al) D_- \ .
\end{equation}
This operator is canonically-equivalent to the harmonic
oscillator, it is defined on the uniform linear lattice in
$x-$space and has the eigenvalues $E_k = k,\ k=0,1,2\ldots$.
However, it does not possess the symmetry $x \rar -x$ unlike the
$\de-$ discretized, canonically-equivalent operator (\ref{e.3.3})
in $y-$space. The operator (\ref{e.4.10}) is non-local, four-point
operator with eigenfunctions
\begin{equation}
\label{e.4.11}
\phi_k (x)= \sum_{\ell=0}^k a_{\ell}\ \om^{\ell \over 2}\
(x+\al)^{(\ell)}\ .
\end{equation}
where $z^{(\ell)}$ is quasi-monomial and $a_{\ell}$ are the
coefficients in the expansion of the Hermite polynomials, $H_k(x)\
=\ \sum_{\ell=0}^k a_{\ell} x^{\ell}$. These eigenfunctions are
closely related to the polynomials which can be called the {\it
modified Hermite polynomials}
\begin{equation}
\label{e.4.12}
 {\hat H}_k(x,\de)=\sum_{\ell=0}^k a_{\ell} x^{(\ell)}\ ,
\end{equation}
and, thus,
\[
\phi_k (x)\ =\ {\hat H}_k(\sqrt{\om} (x+\al),\sqrt{\om} \de)\ .
\]

Summarizing, in this Section we presented two $\de-$ discretized
operators which are canonically-equivalent to the harmonic
oscillator, (\ref{e.3.3}) and (\ref{e.4.10}), which are defined on
the uniform grid in $y-$ and $x-$spaces, respectively. In spite of
the fact that they are represented by different elements of the
Heisenberg-Weyl algebra, they have infinitely-many polynomial
eigenfunctions and are isospectral to the original harmonic
oscillator (1).

\section{\it Dilatation-covariant discretization}

Apart from the translation-covariant discretization given by
$\de$-derivative (see (\ref{e.3.2})) there exists a
dilatation-covariant discretization based on $q$-derivative,
${\cal D}_q$, which is also called the Jackson symbol
\[
{\cal D}_q f(y) = \frac{f(qy) - f(y)}{y(1-q)}\ ,
\]
where $q$ is a complex number. This Section will be devoted to a
brief discussion of the dilatation-covariant discretization or
$q-$discretization. First of all by following the above-mentioned
philosophy a natural question can be posed about an existence of
a quantum canonical conjugate to $q$-derivative -- the operator
which obeys together with ${\cal D}_q$ the commutation relations
(\ref{e.2.1}). Up to our knowledge a definite answer is not found
so far and very likely such an object does not exist in terms of
well-defined operators. However, it is well known that the
derivative ${\cal D}_q$ appears naturally in connection to a
realization of quantum algebras in action on functions in one and
several variables. Thus, it looks reasonable to explore a
generalization of the underlying Heisenberg algebra described
above to the case of the $q-$deformed (quantum) Heisenberg
algebra. Hence, we will study `deformed' quantum systems
possessing a $q-$deformed (quantum) Heisenberg algebra as a
hidden algebra instead of the standard Heisenberg algebra
(\ref{e.2.1}) (see discussion in Introduction).

In order to proceed let us ask first what would happen if in the
expressions (\ref{e.2.6}), (\ref{e4.2}) the operators $a,b$ are
not the generators of the Heisenberg algebra (\ref{e.2.1}) but the
generators of the $q-$deformed Heisenberg algebra
\begin{equation}
\label{e.5.1}
             [a,b]_q \equiv ab  -  qba \ =\ 1\ ,
\end{equation}
where $q$ is a parameter. Following the Theorem proved in
\cite{Turbiner:1994}, one can demonstrate that within the
$q-$deformed Fock space built on the $q-$deformed Heisenberg
algebra (\ref{e.5.1}) there exists the flag $\cal{P}$ of linear
spaces of polynomials in $b$ (see (12)), which is preserved by the
operators (\ref{e.2.6}), (\ref{e4.2}). By a simple calculation one
can find the eigenvalues of the operators (\ref{e.2.6}),
(\ref{e4.2}) in the spectral problem $(*)$
\begin{equation}
\label{e.5.2}
             E_n^{(q)}\ =\ 2\{ n \}\ ,\ n=0,1,\ldots\ ,
\end{equation}
where
\[
 \{ n \}\ =\frac{1-q^n}{1-q}\ ,
\]
is a so-called $q-$number and $\{ n \}\rar n$, if $q \rar 1$. It
is evident that if the parameter $q$ is a real number the spectra
of (\ref{e.2.6}), (\ref{e4.2}) are real.

The algebra (\ref{e.5.1}) has a realization in terms of
$q-$derivative and the operator of multiplication (see, for
example, \cite{Turbiner:1997})
\begin{equation}
\label{e.5.3}
a\ =\ {\cal D}_q ,\quad b\ =\ y \ ,
\end{equation}
with the same $q$ as in (\ref{e.5.1}). This realization has a
property that the vacuum remains the same as in the cases
(\ref{e.3.1})-(\ref{e.3.2}) and without loss of generality it can
be set as $|0>\ =\ 1$. Now we can substitute (\ref{e.5.3}) in (17)
and the following operator emerges
\[
 h_q^{(1)}(y,{\cal D}_q)\ =\ -2\tilde J^0 \tilde J^-
 +2 \tilde J^0 - 2(p+\frac{1}{2}) \tilde J^- \ =
 \]
\begin{equation}
\label{e.5.4}
 -2y {\cal D}_q^2 + 2(y-p-\frac{1}{2}) {\cal D}_q\ ,
\end{equation}
where the generators $\tilde J^0=ba,\ \tilde J^-=a$ have the same
functional form as in (\ref{e.2.3}) but obey the $q-$deformed
commutation relation
\[
[\tilde J^0,\tilde J^-]_{1/q} \ \equiv\ \tilde J^0 \tilde J^- -
\frac{1}{q}\tilde J^0\tilde J^-\ =\ -\tilde J^-\ ,
\]
forming the $q-$deformed Borel subalgebra $b(2)_{q}$ of the
$q-$deformed algebra $sl(2)_q$ \cite{ot} (for discussion see
\cite{Turbiner:1994}). In this case the operators (\ref{e.2.6}),
(\ref{e4.2}) are the {\it $sl(2)_q$-exactly-solvable operators}.
Moreover, the operator (\ref{e.2.6}) (as well as (\ref{e4.2})) can
be called the {\it $q-$deformed harmonic oscillator Hamiltonian in
Fock space possessing $sl(2)_q$ hidden algebra}.

The operator $h_q(y, D_q)$ is a non-local, three-point, discrete,
dilatation-covariant operator defined on exponential lattice. It
is illustrated by Fig.3.

\vskip .5cm
\noindent
\unitlength.8pt
\begin{picture}(400,50)(-10,-20)
 \linethickness{1.2pt}
 \put(60,10){\line(1,0){310}}
 \put(120,10){\circle*{5}}
 \put(180,10){\circle*{5}}
 \put(300,10){\circle*{5}}
 \put(110,-15){$\phi(y)$}
 \put(170,-15){$\phi(qy)$}
 \put(280,-15){$\phi(q^2y)$}
\end{picture}
\begin{center}
Fig. 3. \ Graphical representation of the operator (\ref{e.5.4})
\end{center}

\vskip .5cm The operator (\ref{e.5.4}) can be called {\it
the algebraic form  of the Hamiltonian of the $q-$discretized
harmonic oscillator}.

The spectral problem for the operator (\ref{e.5.4}) has a form
\[
-\frac{2}{yq(q-1)^2} \phi(q^2y) +\Bigg[
\frac{2+q+q^2 - 2 pq(1-q)}{q(q-1)^2}\frac{1}{y}
+\frac{2}{1-q}\Bigg]
\phi(qy)-
\]
\begin{equation}
\label{e.5.5}
 \Bigg[
 \frac{1+ q-2p(1-q)}{y(q-1)^2}+\frac{2}{1-q}
 \Bigg]\phi(y)\ =\ E^{(q)} \phi(y) \ .
\end{equation}
or, the r.h.s. can be taken as
\begin{equation}
\label{e.5.6}
 \ =\ E^{(q)} \phi(qy) \ .
\end{equation}
or as
\begin{equation}
\label{e.5.7}
 \ =\ E^{(q)} \phi(q^2y) \ .
\end{equation}
Usually, the spectral problem for $q-$discrete operators is
defined with (\ref{e.5.6}) as the r.h.s. (see, for example,
\cite{Exton:1983}). It assumes that the middle point in Fig.3
remains fixed under dilatation. Introducing the new variable
$\tilde y = qy$, it can be seen explicitly.

If in the case (\ref{e.5.5}) the eigenvalues are given by
(\ref{e.5.2}) while for (\ref{e.5.6}), (\ref{e.5.7}) the
eigenvalues are equal to
\begin{equation}
\label{e.5.8}
   E_n^{(q)}\ =\ -2\{ -n \}\ ,\ n=0,1,\ldots
\end{equation}
\begin{equation}
\label{e.5.9}
   E_n^{(q)}\ =\ 2 q^{-2n} \{ n \}\ ,\ n=0,1,\ldots
\end{equation}
correspondingly. In the limit $q \rar 1$ all three expressions
coincide corresponding to the original harmonic oscillator
spectrum. The spectral problems (\ref{e.5.5})--(\ref{e.5.7}) can
be considered as a possible definition of a $q-$deformed harmonic
oscillator. In the literature it is well-known many other
definitions of the $q-$deformed harmonic oscillator (see for
example, \cite{Zachos}, \cite{AFW} and references therein,
\cite{NV})
 \footnote{Usually, these deformations are done by a direct
 discretization of the original Hamiltonian (1). Most of all
 are based on a discretization of the Infeld-Hall
 factorization representation of (1)}.
Such a situation reflects an ambiguity of making a
$q-$deformation
 \footnote{For example, any term in non-deformed expression
 can be modified by multipliers of the
 type $q^a$ and even some extra terms can be added with vanishing
 coefficients in the limit $q\rar 1$ like $(1-q)^b, b>0$}
as well as absence of clear physical criteria, which can remove or
reduce this ambiguity. For instance, in the literature it is
exploited three different types of the $q-$Laguerre polynomials
(see, for example, an excellent review \cite{Koekoek:1994}), but
it is not clear why other possible $q-$deformations of Laguerre
polynomials are not studied.

Isospectrality of (17) and (28) is preserved by the
$q-$deformation. Substitution of (\ref{e.5.3}) in (\ref{e4.2})
gives a slight modification of the expressions (\ref{e.5.4}).
Unlike translation-covariant case it does not lead to a change of
the nature of non-locality changing the number of points in the
operator (\ref{e.5.4}) as it is happened for the operators
(\ref{e.3.3}) and (\ref{e.4.3}). The $q-$deformation of another
algebraic form of the harmonic oscillator (\ref{e.4.8}) in Fock
space
\begin{equation}
\label{e.5.10}
 h_q^{(2)}(b,a)\ =
 \ -\frac{1}{2\om}{\tilde J}^- {\tilde J}^- + {\tilde J}^0  \ ,
\end{equation}
(cf.(\ref{e.4.9})) takes in terms of the $q-$derivative the
following form
\begin{equation}
\label{e.5.11}
 h_q^{(2)}(x,{\cal D}_q)\ =\ - \frac{1}{2\om}{\cal D}_q^2\
+\ x {\cal D}_q\ ,
\end{equation}
(cf.(\ref{e.5.4})). It should be mentioned that the operators
(\ref{e.5.4}) and (\ref{e.5.11}) are defined on the essentially
different lattices, which are exponential in $x-$ and
$y-$variables, respectively.

Similar to what was done for canonical transformations it seems natural
to introduce a notion of $q-$deformed canonical transformations, when
two $q-$deformed systems are $q-$canonically equivalent if they
can be connected through the $q-$deformed canonical transformation
(see below). However, we were unable to find well-defined,
non-trivial realization of the algebra (\ref{e.5.1}) other than
(\ref{e.5.2}) which, for instance, would be similar to the
realization (\ref{e.3.3}) for (\ref{e.2.1}) (see discussion in
\cite{Turbiner:1997}) .

\section{\it Anharmonic oscillator (perturbation theory)}

Anharmonic oscillator is one of the most important
non-exactly-solvable problems of quantum mechanics. One of the
simplest and the most popular examples is given by the Hamiltonian
\begin{equation}
\label{e.6.1}
        {\cal H}^{(aho)}= -\frac{1}{2} \frac{\pa^{2}}{\pa x^{2}} +
        \frac{\om^2}{2} x^{2} + g x^{2n} \ ,\ n=2,3,4,\ldots
\end{equation}
where $g$ is a coupling constant\footnote{So far, the most
comprehensive study of the anharmonic oscillator (\ref{e.6.1}) was
carried out by Bender-Wu in the classical paper \cite{BW} (see
also \cite{Turbiner:1984}) }. Making first the gauge rotation of
(\ref{e.6.1}) with the gauge factor (\ref{e1.4}) we get
\[
 h^{(aho)}(y,\pa_y)\ =\ \frac{1}{\om}\ (\Psi_0^{(p)}(x))^{-1}
     {\cal H}^{(aho)} \Psi_0^{(p)}(x)\mid_{y=\om x^2}
\]
\begin{equation}
\label{e.6.2}
 =\ -2y\pa_y^2 + 2(y-p-\frac{1}{2})\pa_y\ +\ \la y^n\ ,
\end{equation}
(cf.(\ref{e1.5})), where $p=0,1$ and has a meaning of parity and
$\la= g\om^{-(n+1)}$. By the reasons which will be clear later the
operator (\ref{e.6.2}) can be called {\it algebraic form of the
anharmonic oscillator Hamiltonian}. Replacing in (\ref{e.6.2}) the
derivative and the coordinate by the elements of the Heisenberg
algebra (\ref{e.2.1}), $\pa_y \rar a, y \rar b$ we will arrive at
the element of the Heisenberg-Weyl algebra
\begin{equation}
\label{e.6.3}
 h^{(aho)}(b,a)\ =\ -2ba^2 + 2(b-p-\frac{1}{2})a\ +\ \la b^n\ ,
\end{equation}
which can be called the {\it anharmonic oscillator Hamiltonian in
the Fock space}. Now we can fix $n=2$ and we will consider this
particular case of the {\it quartic anharmonic oscillator} as the
major example in further consideration.

Let us find a finite-difference operator which is canonically
equivalent to (\ref{e.6.1}). In order to do it we simply
substitute the realization (\ref{e.3.2}) of the Heisenberg algebra
to the operator (\ref{e.6.3})
\[
  h_{\de}^{(aho)} (y, D_{\pm})\
 =\ -\frac{2}{\de}[y+\al+\de(p+\frac{1}{2})]D_+\ +
 \ 2[(1 +\frac{1} {\de})(y+\al)\ -
\]
\begin{equation}
\label{e.6.4}
 \frac{\de \la}{2} (y+\al)^{(2)}]D_- + \frac{\de^2
 \la}{2} (y+\al)^{(2)} D_- D_- + \frac{\la}{2} (y+\al)^{(2)}.
\end{equation}
(see Fig.4). It is quite amazing that the perturbation $\la b^2$
leads solely to an addition of one more point (marked by
$\diamond$) to the harmonic oscillator operator. A model
characterized by the operator (\ref{e.6.4}) can be called a {\it
finite-difference anharmonic oscillator}.
 \vskip .5cm
 \noindent
 \unitlength.8pt
\begin{picture}(400,50)(-10,-20)
 \linethickness{1.2pt}
 \put(50,10){\line(1,0){400}}
 \put(150,6){$\diamond$}
 \put(220,10){\circle*{5}}
 \put(290,10){\circle*{5}}
 \put(360,10){\circle*{5}}
 \put(110,-15){$\phi(y-2\de)$}
 \put(200,-15){$\phi(y-\de)$}
 \put(280,-15){$\phi(y)$}
 \put(340,-15){$\phi(y+\de)$}
\end{picture}
\begin{center}
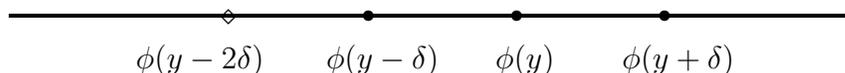
Fig. 4. \ Graphical representation of the problem (\ref{e.6.4})
\end{center}
It is non-local, four-point, finite-difference operator (for
comparison see Fig.1 and Fig.2 with $\de \rar -\de$).

The corresponding spectral problem $(*)$ looks as follows
\[
-\frac{2}{\de^2}\big[y+\al+\de(p+\frac{1}{2})\big]\phi(y+\de)
+\ \frac{2}{\de}\big[(1+\frac{2}{\de})y+\al+p+\frac{1}{2}\big]
\phi(y)\ -
\]
\begin{equation}
\label{e.6.5}
 \frac{2}{\de}(1+\frac{1}{\de})(y+\al)\phi(y-\de)\ +\
 \la (y+\al)^{(2)}\phi(y-2\de) \ =\ E \phi(y) \ .
\end{equation}
Their eigenvalues coincides with those of the anharmonic
oscillator (\ref{e.6.1})--(\ref{e.6.3}).

We have to emphasize that a presence of the anharmonic term
changes the nature of non-locality of the harmonic oscillator
leading to appearance an extra point. In particular case $\de=-1$
the number of points is reduced to three, however, the lattice
becomes non-uniform.

In similar way one can construct a $q-$deformed anharmonic
oscillator taking the operator (\ref{e.6.3}) as an element of the
$q-$Fock space. Substituting the realization (\ref{e.5.3}) of the
$q-$deformed Heisenberg algebra into (\ref{e.6.3}) we get
\begin{equation}
\label{e.6.6}
 h_q^{(aho)}(y,{\cal D}_q)\ =\ -2y {\cal D}_q^2
 +2(y-p-\frac{1}{2}) {\cal D}_q + \la y^2\ ,
\end{equation}
(cf. (\ref{e.5.4})) which three-point, non-local operator (see,
for example, Fig.3).

The spectral problem $(*)$ for the operator (\ref{e.6.6}) has a
form
\[
 -2\frac{\phi(q^2y)}{yq(q-1)^2} +2
 \frac{1+q+(y-p-\frac{1}{2})q(1-q)}{yq(q-1)^2}\phi(qy)-
\]
\begin{equation}
\label{e.6.7}
 \frac{2-(2p+1)(1-q)+2y(1-q)-\la y^2(q-1)^2}{y(q-1)^2}\phi(y)
 \ =\ E^{(q)} \phi(y) \ .
\end{equation}
(cf.(\ref{e.5.4})). Certainly, the r.h.s. in (\ref{e.6.7}) can
vary being equal to either (\ref{e.5.6}), or (\ref{e.5.7}). It
reflects a fact that in the case of the $q-$Fock space the
spectral problem $(*)$ can be modified in one way
\begin{equation*}
 L(b,a)\phi (b)|0>\ =\ \la \phi (qb) |0>\ ,\qquad\qquad\qquad (**)
\end{equation*}
or another
\begin{equation*}
L(b,a)\phi (b)|0>\ =\ \la \phi (q^2b) |0>\ .\qquad\qquad\qquad
(***)
\end{equation*}
If we introduce the multiplication operator $T_q f(x)=f(qx)$ the
problems $(**)$ or $(***)$ correspond to an appearance of
non-trivial operator weight factors in r.h.s.
\begin{equation*}
L(b,a)\phi (b)|0>\ =\ \la T_q \phi (b) |0> \ ,
\end{equation*}
or
\begin{equation*}
L(b,a)\phi (b)|0>\ =\ \la T^2_q \phi (b) |0> \ ,
\end{equation*}
respectively.

It can be easily shown that the operator (\ref{e.6.3}) describing
the anharmonic oscillator does not belong to the class of the
$sl(2)-$exactly-solvable operators. Hence their eigenfunctions are
not polynomials. However, it was proved in \cite{Turbiner:1999}
that in framework of some perturbative approach (see below) as a
consequence of the fact that the perturbation $b^2 \in {\cal
P}_{2}(b)$ (see (\ref{e.2.5})) the perturbation theory in powers
of the parameter $\la$ is algebraic one: any correction to any
eigenfunction is a finite-order polynomial and hence can be found
by algebraic means. In fact, such a perturbative approach provides
a certain regular way to define an object in the Heisenberg-Weyl
algebra which can be called the Hamiltonian of the quartic
anharmonic oscillator. From practical point of view such a
perturbation theory has very important feature: once being
developed for the operator (\ref{e.6.3}) it gives a unique
possibility to construct {\it simultaneously (!)} the perturbation
theory for the operators (\ref{e.6.1}), (\ref{e.6.5}),
(\ref{e.6.7}).

Let us approach to a construction of the above-mentioned
perturbation theory. Following a standard prescription we take the
spectral problem $(*)$ with $$L(b,a)=h_0+\la h_1$$ and develop a
perturbation theory in powers of $\la$ searching for corrections
in a form
\begin{equation}
\label{pt.1}
 \phi = \sum \la^n \phi_n \ ,\ E = \sum \la^n E_n \ .
\end{equation}
Collecting the terms of the order $\la^n$ it is easy to derive an
equation for the $n$th correction
\begin{equation}
\label{pt.2}
 (h_0 - E_0) \phi_n = \sum_{i=1}^{n} E_i \phi_{n-i} - h_1
 \phi_{n-1}\ .
\end{equation}
A remarkable feature of this form of perturbation theory a
possibility to study a single state separately, without touching
other states as it was the case of the Rayleigh-Schroedinger form
of perturbation theory.

Now we take as the unperturbed operator $h_0$ the Hamiltonian of
the harmonic oscillator in the Fock space (\ref{e.2.6}) and
consider as the perturbation $h_1=b^2$. As it was mentioned
already, in this case the $n$th correction $\phi_n$ to
eigenfunction should be polynomials in $b$. As an instructive
example let us study the ground state of anharmonic oscillator.
The ground state of the unperturbed problem (\ref{e.2.6}) is
characterized by
\begin{equation}
\label{pt.3}
 \phi_0=1\ ,\ E_0=1\ .
\end{equation}
By solving the equation (\ref{pt.2}) after simple calculations we
can get the explicit form for the first several corrections, for
example,
\[
 -\phi_1\ =\ \frac{b^2}{2\{ 2\}}+\frac{(3+2p)}{4}b\ ,
 \quad E_1=\frac{(1+2p)(3+2p)}{4}\ ,
\]
and
\[
 \phi_2\ =\ \frac{1}{4}\bigg[\frac{1}{\{2\}\{4\}}b^4\ +
 \frac{2q^2+5q+6+2p(q+2)}{2\{2\}\{3\}}b^3\ +
\]
\[
 \frac{4q^2+12q+15+ 8p(q+2)+4p^2}{4\{2\}}b^2\ +
 \frac{(3+2p)[q^2+3q+3+2p(q+1)]}{2}b\bigg]\ ,
\]
\begin{equation}
\label{pt.4}
  E_2\ =\ -\frac{(1+2p)(3+2p)}{8}\big[q^2+3q+3+2p(q+1)\big]\ .
\end{equation}
Without any difficulties one can find several next corrections.
However, it becomes evident very quickly that the complexity of
calculations is growing very fast with a number of correction.
Using different realizations of the $(q)-$Heisenberg algebra one
can calculate perturbative corrections to the various forms of
anharmonically-perturbed harmonic oscillator (differential,
finite-difference, discrete).

\bigskip

\begin{itemize}

\item[I.] $q=1,\ b=y$.\\

\noindent
This case corresponds to the coordinate-momentum realization of
the Heisenberg algebra (\ref{e.3.1}) and vacuum definition
$|0>=1$. It leads to a standard anharmonic oscillator
(\ref{e.6.1}) and the corrections are:
\[
 -\phi_1\ =\ \frac{y^2}{4}+\frac{(3+2p)}{4}y\ ,
\]
and
\[
 \phi_2\ =\ \frac{1}{4}\bigg[\frac{y^4}{8}+
 \frac{11+6p}{12}y^3\ +\ \frac{31+24p+4p^2}{8}y^2\ +
\]
\begin{equation}
\label{pt.5}
 \frac{(3+2p)(7+4p)}{2}y\bigg]\ ,
 \ E_2\ =\ -\frac{(1+2p)(3+2p)(7+4p)}{8}\ ,
\end{equation}
where $E_1$ is the same as in (\ref{pt.4}).

Given form of the perturbation theory coincides to the so-called
`$F-$functions method' developed by Dalgarno (see discussion in
\cite{Turbiner:1984}), which in fact was realized for the case of
the ahnarmonic oscillator (\ref{e.6.1}) in \cite{BW}. It is easy
to check that the corrections (\ref{pt.5}) coincide to those
calculated in text-books (see for example \cite{llqm}).\\

\item[II.] $q=1,\ b=(y+\al)(1-\de{\cal D}_-)$ (see (\ref{e.3.2}))\\

\noindent
This case corresponds to the translation-covariant discretization
and perturbed finite-difference harmonic oscillator (\ref{e.3.3})
(see (\ref{e.6.4})):
\[
 -\phi_1\ =\ \frac{{\tilde y}^{(2)}}{4}+\frac{(3+2p)}{4}{\tilde y}\ ,
\]
and
\[
 \phi_2\ =\ \frac{1}{4}\bigg[\frac{{\tilde y}^{(4)}}{8}+
 \frac{11+6p}{12}{\tilde y}^{(3)}\ +\ \frac{31+24p+4p^2}{8}
 {\tilde y}^{(2)}\ +
\]
\begin{equation}
\label{pt.6}
 \frac{(3+2p)(7+4p)}{2}{\tilde y}\bigg]\ ,
\end{equation}
where $E_1$ is the same as in (\ref{pt.4}) and $E_2$ is the same
as in (\ref{pt.5}). Here ${\tilde y}^{(n+1)} = (y+\al)^{(n+1)} =
(y+\al)(y+\al-\de)\ldots (y+\al-n\de)$ is quasi-monomial.\\

  \item[III.] $q \neq 1,\ b=y$.\\

\noindent
This case corresponds to the perturbed $q-$harmonic oscillator
(\ref{e.6.6})-(\ref{e.6.7}) and corrections are equal to
\[
 -\phi_1\ =\ \frac{y^2}{2\{ 2\}}+\frac{(3+2p)}{4}y\ ,
 \quad E_1=\frac{(1+2p)(3+2p)}{4}
\]
and
\[
 \phi_2\ =\ \frac{1}{4}\bigg[\frac{1}{\{2\}\{4\}}y^4+
 \frac{2q^2+5q+6+2p(q+2)}{2\{2\}\{3\}}y^3\ +
\]
\[
 \frac{4q^2+12q+15+ 8p(q+2)+4p^2}{4\{2\}}y^2\ +
 \frac{(3+2p)[q^2+3q+3+2p(q+1)]}{2}y\bigg]\ ,
\]
\begin{equation}
\label{pt.7}
  E_2\ =\ -\frac{(1+2p)(3+2p)}{8}\big[q^2+3q+3+2p(q+1)\big]\ .
\end{equation}

\end{itemize}

It is quite interesting to see how above-mentioned results will be
modified if instead of the spectral problem $(*)$ the spectral
problem $(**)$ is considered. The equation (\ref{pt.2}) for the
$n$th correction becomes
\begin{equation}
\label{pt.8}
 (h_0 - E_0 T_q) \phi_n = \sum_{i=1}^{n} E_i T_q \phi_{n-i} - h_1
 \phi_{n-1}\ ,
\end{equation}
(cf. (\ref{pt.2})). The ground state of the unperturbed problem
(\ref{pt.3}) is unchanged. The first correction is also unchanged
while the second correction now takes a modified form
\[
 \phi_2\ =\ \frac{1}{4}\bigg[\frac{1}{\{2\}\{4\}}b^4+
 \frac{2q^2+5q+6+2p(q+2)}{2\{2\}\{3\}}b^3\ +
\]
\[
 \frac{4q^2+12q+15+ 8p(q+2)+4p^2}{4\{2\}}b^2\ +
 \frac{(3+2p)[q^2+3q+3+2p(q+1)]}{2}b\bigg]\ ,
\]
\begin{equation}
\label{pt.9}
 E_2\ =\ -\frac{(1+2p)(3+2p)}{64}\big[65-9q^2\ +\ 8p(7-3q^2)\ +
 12p^2 (1-q^2)\big]\ .
\end{equation}

\section{\it Anharmonic oscillator (quasi-exactly-solvable model)}

Among one-dimensional Schroedinger equations there is some class
of problems possessing a certain outstanding property - first
several eigenstates can be found explicitly, by algebraic means.
Such problems are called {\it quasi-exactly-solvable} \cite{tu}.
Among ten known families of one-dimensional quasi-exactly-solvable
potentials \cite{Turbiner:1988} there exists one, which can be
treated as an anharmonic oscillator and its Hamiltonian is
\begin{equation}
\label{qes.1}
 {\cal H}^{(qes)}= - \frac{1}{2} \frac{\pa^{2}}{\pa x^{2}} +
 [\frac{\om^2}{2}-(2n+\frac{3}{2}+p)g] x^2 + g\om x^4+
 \frac{g^2}{2} x^6 \ .
\end{equation}
Here $g$ is a coupling constant and $x \in (-\infty,\infty)$.
Parameter $p=0,1$ has meaning of parity. The first $(n+1)$
eigenfunctions of parity $p$ (but not others) are of the form
\begin{equation}
\label{qes.2}
 \Psi_n^{(p)}(x)= x^p P_n(\om x^2) e^{-\om x^2/2 - g x^4/4}
\end{equation}
where $P_n(y)$ is a polynomial of degree $n$.

Making first the gauge rotation of (\ref{e.6.1}) with the gauge
factor (\ref{qes.2}) at $n=0$ we get
\[
h^{(qes)}(y,\pa_y)\ =\ \frac{1}{\om}\ (\Psi_0^{(p)}(x))^{-1}
{\cal H}^{(qes)}\Psi_0^{(p)}(x)\mid_{y=\om x^2}
\]
\begin{equation}
\label{qes.3}
=\ -2y\pa_y^2 +
2(\la y^2+ y-p-\frac{1}{2})\pa_y\ -\ 2\la n y\ ,
\end{equation}
(cf. (\ref{e1.5}),(\ref{e.6.2})), where $\la=g \om^{-2}$ and
constant terms are dropped out. The spectral problem for
(\ref{qes.3}) is defined on the half-line, $y \in [0,\infty)$. The
first $(n+1)$ eigenfunctions of (\ref{qes.3}) are some polynomials
of the degree $n$, $P_n(y)$ (cf.(\ref{qes.2})) possessing
$k=0,1,\ldots, n$ real zeroes inside of the interval $y \in
[0,\infty)$.

Replacing in (\ref{qes.3}) the derivative and the coordinate by
the elements of the Heisenberg algebra (\ref{e.2.1}): $\pa_y \rar
a, y \rar b$ we arrive at the element of the Heisenberg-Weyl
algebra
\begin{equation}
\label{qes.4}
h^{(qes)}(b,a)\ =\ -2ba^2 + 2(\la b^2+ b-p-\frac{1}{2})a\
-\ 2\la n b
\ ,
\end{equation}
(cf. (\ref{e.2.6}),(\ref{e4.2}),(\ref{e.6.3})). As in previous
consideration the operator $h^{(qes)}(b,a)$ can be treated as an
element of the Fock space as well as an element of the $q-$Fock
space. This operator can be called the {\it Fock space Hamiltonian
of the anharmonic quasi-exactly-solvable oscillator}. The operator
$h^{(qes)}(b,a)$ in the Fock space is
$sl_2-$quasi-exactly-solvable operator. It can be rewritten in
terms of the generators of $sl_2-$algebra (\ref{e.2.3})
\begin{equation}
\label{qes.4.1}
h^{(qes)}(b,a)= -2 J^0_n J^-_n + 2\la J^+_n + 2 J^0_n -
(n+1+2p)J^-_n\ ,
\end{equation}
where as always the constant terms are dropped off.

However, it is easy to check that in the $q-$Fock space the
operator $h^{(qes)}(b,a)$ is {\it not}
$sl_{2q}-$quasi-exactly-solvable operator, since it can not be
rewritten in terms of the generators of
$sl_{2q}-$algebra
 \footnote{In the case of the $sl_{2q}-$algebra
 the major modification of (\ref{e.2.3}) comes for the
 positive-root generator
\[
J^+_n = b^2 a - n b \rar {\tilde J}^+_n = b^2 a -\{n\} b\ ,
\]
 while the Cartan generator
\[
 J^0_n
 = b a - \frac{n}{2} \rar {\tilde J}^0_n = b a
 -\frac{\{n\}\{n+1\}}{\{2n+2\}}\ ,
\]
 where $\{n\}$ is the $q-$number (see (31)) and $J^-$ remains
 unchanged; for details, see \cite{Turbiner:1997}}.
This operator needs a slight modification of the last term in
order to become the $sl_{2q}-$quasi-exactly-solvable:
\begin{equation}
\label{qes.4.2}
 h^{(qes)}(b,a)\ =\ -2ba^2 + 2(\la b^2+ b-p-\frac{1}{2})a\
 -\ 2\la \{n\} b\ = \
\end{equation}
\[
 -2 {\tilde J}^0_n {\tilde J}^-_n + 2\la {\tilde J}^+_n +
 2 {\tilde J}^0_n - \bigg(2\frac{\{n\}\{n+1\}}{\{2n+2\}}+1+2p\bigg)
  {\tilde J}^-_n \ ,
\]
(cf. (\ref{qes.4})), where $\{n\}$ is the $q-$number (see (31))
and the constant terms in the second expression are dropped off.

The first $(n+1)$ eigenfunctions of (\ref{qes.4.2}) are some
polynomials in $b$ of degree $n$, $P_n(b)$. If in the case $n=0$
the first eigenfunctions for the spectral problems $(*)$ and
$(**)$ coincide and equal to:
\[
 \phi^{(0)}=1\ ,\ E^{(0)}=0\ ,
\]
while for the case of $n=1$ they are already different. Namely,
for the spectral problem $(*)$,
\[
 \phi^{(1)}_{\pm}\ =\ b\ -\ \frac{1\mp \sqrt{1+4\la (1+2p)}}
 {4\la}\ ,
 \]
\begin{equation}
\label{qes.5}
 E^{(1)}_{\pm}=\frac{1\mp \sqrt{1+4\la (1+2p)}}{2}\ ,
\end{equation}
where sub-indices $(+)$ and $(-)$ are assigned to the ground and
the first excited state, respectively. As for the spectral problem
$(**)$ the formulas (\ref{qes.5}) are modified
\[
 \phi^{(1)}_{\pm}\ =\ b\ -\ \frac{1\mp \sqrt{1+4\la q
 (1+2p)}}{4\la}\ ,
 \]
\begin{equation}
\label{qes.6}
\ E^{(1)}_{\pm}=\frac{1\mp \sqrt{1+4\la q (1+2p)}}{2q}\ .
\end{equation}
Of course, when $q=1$ the formulas (\ref{qes.5}) and
(\ref{qes.6}) coincide.

Now we can find a finite-difference operator canonically
equivalent to (\ref{qes.3}). In order to do it we put $q=1$ in
(\ref{qes.3}) and substitute the realization (\ref{e.3.2}) of the
Heisenberg algebra into the operator (\ref{qes.4})
\[
 h_{\de}^{(qes)} (y, D_{\pm})\ =
\]
\[
 -\frac{2}{\de}[y+\al+\de(p+\frac{1}{2})]D_+\ +
 2[(1 +\frac{1}{\de})(y+\al)+\la(y+\al)(y+\al+\de(n-1))]D_-
\]
\begin{equation}
\label{qes.7}
 -2\la \de (y+\al)^{(2)} D_- D_-\ -\ 2\la n (y+\al) \ ,
\end{equation}
(cf. (\ref{e.6.4})). It turns out that this operator is four-point
finite-difference operator (see Fig.4). It is quite amazing that
independently on $n$ the perturbation $2\la b(b a-n)$ to the
harmonic oscillator operator (see Fig.1) leads solely to an
appearance of one extra point (marked by $\diamond$ in Fig.4)
similarly to (\ref{e.6.4}). A model characterized by the operator
(\ref{qes.7}) can be called a {\it finite-difference
quasi-exactly-solvable anharmonic oscillator}.

Without loss of generality we place $\al=0$ in (\ref{qes.7}) and
the corresponding spectral problem $(*)$ for (\ref{qes.7}) looks
like
\[
 -\frac{2}{\de}\big[\frac{y}{\de}+p+\frac{1}{2}\big] \phi(y+\de) +\
 \frac{2}{\de}\big[(1+\frac{2}{\de})y+p+\frac{1}{2}\big] \phi(y)
\]
\[
 -\ \frac{2}{\de}y \Big[(1+\frac{1}{\de})-\la (y-\de(n+1))\Big]
 \phi(y-\de)
\]
\begin{equation}
\label{qes.8}
 -\ 2\la \frac{y^{(2)}}{\de}\phi(y-2\de) \ =\ E\phi(y) \ .
\end{equation}
It is natural to assume that the spectral problem (\ref{qes.8}) is
defined in $y \in [0,\infty)$. Their first $(n+1)$ eigenvalues
coincide with those of the anharmonic oscillator Hamiltonian
(\ref{qes.1}) as well as the operators
(\ref{qes.3}),(\ref{qes.4}). Their eigenfunctions remain
polynomials but modified -- each monomial should be replaced by
quasi-monomial, $y^k \rar y^{(k)}$. For instance, the formulas
(\ref{qes.5})--(\ref{qes.6}) become
\[
 \phi^{(1)}_{\pm}\ =\ y\ -\ \frac{1\mp \sqrt{1+4\la (1+2p)}}
 {4\la}\ ,
\]
\begin{equation}
\label{qes.9}
 E^{(1)}_{\pm}=\frac{1\mp \sqrt{1+4\la (1+2p)}}{2}\ .
\end{equation}
It is worth to note that for physically relevant parameters $g
\geq 0$ and $q \geq 0$
 \footnote{In this case $q$ has the meaning the parameter of
 dilatation}
the eigenfunctions (\ref{qes.5}), (\ref{qes.9}) in domain $y\geq
0$ have no nodes for ground state and have the only one node for
the first excited state like an analogue of the oscillation
(Sturm) theorem holds (see, for example, \cite{llqm}). Likely, an
agreement with Sturm theorem will hold for higher excited states
but we were unable to prove it in full generality.

\section{Conclusion}

We introduced a simple-minded notion of canonical equivalence in
quantum mechanics. Then we showed that for canonically-equivalent
systems the spectra remain the same and constructed discrete
systems, which are canonically-equivalent to the harmonic and a
certain type of anharmonic oscillators -- the quartic oscillator
and the quasi-exactly-solvable oscillator. In general, these
discrete systems are defined on non-linear lattices.

We restricted our studies to the operators in the algebraic form
possessing the polynomial eigenfunctions. An important question to
ask is what would happen if the operator in hand possesses an
algebraic form but non-polynomial eigenfunctions. Perhaps, the
most explicit instructive example is given by the original
Hamiltonian (1) of the harmonic oscillator. Following our
philosophy it can be rewritten as the element of the Fock space
\begin{equation}
\label{fin.1}
 {\cal H}\ =\ -\frac{1}{2}a^2 + \frac{\om^2}{2}b^2\ ,
\end{equation}
where its ground state eigenfunction can be written formally as
\begin{equation}
\label{fin.2}
 \Psi_0(b) |0>\ =\ \sum_{n=0}^{\infty}\frac{(-\om)^n}{2^n n!}
 b^{2n} |0>\ .
\end{equation}
Substitution of (19) with $|0>=1$ into (78) leads to an infinite
series of the form
\begin{equation}
\label{fin.3}
{\tilde \Psi}_0(x)\ =\ \sum_{n=0}^{\infty}\frac{(-\om)^n}{2^n n!}
 x^{(2n)} \ ,
\end{equation}
which has zero radius of convergence in $x$ for any $\de \neq 0$
 \footnote{This type of expansion on the basic set of quasi-monomials
 is known in literature as the Newton series. For discussion see, for
 example, \cite{finite}}.
Although the Taylor expansion in powers of $x$ for the original
eigenfunction (4) at $p=0$ had infinite radius of convergence. So,
the radius of convergence has delta-function behaviour. Perhaps,
one of possible ways to remedy this drawback is to gauge-rotate a
Hamiltonian with non-polynomial eigenfunctions using a
semi-classical or somehow modified semi-classical wavefunction as
a gauge factor requiring an appearance of an algebraic form of the
final operator.

\bigskip
 {\it Acknowledgement}

\smallskip

I want to express my deep gratitude to M.~Shifman, M.~Shubin,
E.~Shuryak, Yu.F.~Smirnov and N.~Vasilevsky for their sincere
interest to the work and valuable discussions, and especially to
B.~Julia for the conversation, which after some time turned out to
be very inspiring.

\newpage
\def\href#1#2{#2}

\begingroup\raggedright
\endgroup

\end{document}